\documentclass[pdflatex,sn-mathphys-num]{sn-jnl}% Math and Physical Sciences Numbered Reference Style
\usepackage{graphicx}%
\usepackage{multirow}%
\usepackage{amsmath,amssymb,amsfonts}%
\usepackage{amsthm}%
\usepackage{mathrsfs}%
\usepackage[title]{appendix}%
\usepackage{xcolor}%
\usepackage{textcomp}%
\usepackage{manyfoot}%
\usepackage{booktabs}%
\usepackage{algorithm}%
\usepackage{algorithmicx}%
\usepackage{algpseudocode}%
\usepackage{listings}%
\usepackage{lmodern} 
\usepackage{microtype}
\usepackage{float}
\usepackage{placeins}
\usepackage{enumitem}
\usepackage{caption}
\usepackage{tabularx}
\usepackage{array}
\usepackage{threeparttable}
\usepackage{lscape}
\usepackage{rotating}
\usepackage{pdflscape}
\usepackage{geometry}
\usepackage{subcaption}
\makeatletter
\gdef\orcidlogo{} % don't load the EPS
\renewcommand{\orcid}[1]{} % disable clickable ORCID
\makeatother

\newlength{\portraittextwidth}
\renewcommand{\tablename}{Table}

\makeatletter
\renewcommand\@biblabel[1]{}
\makeatother

\theoremstyle{thmstyleone}%
\theoremstyle{thmstyletwo}%

\theoremstyle{thmstylethree}%
\usepackage{tcolorbox}
\usepackage{draftwatermark}
\usepackage{tikz}
\usepackage{eso-pic}
\usepackage{fancyhdr}% allows background layers
\usepackage{fancyhdr}

\fancypagestyle{mystyle}{
    \fancyhf{} % clear header & footer

    \fancyhead[L]{\small Leicester’s Tale: xG Modelling for EPL 2015/16}
    \fancyhead[R]{\small Preprint} % right-side note % left header
    \fancyfoot[C]{\thepage} % page number centered
}
\makeatletter
\renewcommand{\keywordname}{\hspace*{-\leftmargin}Keywords} % left-align Springer heading
\makeatother

\begin{document}

% ---- Watermark only on first page ----
\AddToShipoutPictureBG*{
    \begin{tikzpicture}[remember picture,overlay]
        \node[rotate=45, text opacity=0.2, scale=4] at (current page.center) {PREPRINT};
    \end{tikzpicture}
}

\fancypagestyle{firstpagefooter}{
    \fancyhf{} % clear everything
    \renewcommand{\headrulewidth}{0pt}
    \renewcommand{\footrulewidth}{0.4pt}

    % one-line, left-aligned footer (no wrapping)
    \fancyfoot[L]{%
        \makebox[\linewidth][l]{%
            \normalsize *Corresponding author e-mail: \mbox{\texttt{sheikhbadaruddin.tahir@phd.units.it}}
        }%
    }
}
% \title{Leicester’s Tale: Another Perspective on the EPL 2015/16 Through Expected Goals (xG) Modelling}
\title{
\makebox[\textwidth][c]{%
    \parbox{1.4\textwidth}{%
        \centering
        {\bfseries\LARGE 
        Leicester's Tale: Another Perspective on the EPL 2015/16 Through Expected Goals (xG) Modelling}%
    }%
}
}
% \author*{\fnm{\textbf {Sheikh Badar Ud Din}} \sur{\textbf {Tahir}}}
% \author{\fnm{\textbf {Leonardo}} \sur{\textbf {Egidi}}}
% \author{\fnm{\textbf {Nicola}} \sur{\textbf {Torelli}}}
\author{\textbf{Sheikh Badar Ud Din Tahir}$^{*}$}
\author{\textbf{Leonardo Egidi}}
\author{\textbf{Nicola Torelli}}
\affil{\orgdiv{Department of Economics, Business, Mathematics and Statistics ``Bruno de Finetti''}, 
\orgname{University of Trieste},
\orgaddress{\city{Trieste}, \postcode{34127}, \country{Italy}}}

% ---- ABSTRACT MUST BE HERE ----
\abstract{
\begin{center}
\begin{minipage}{1.1\textwidth}

Probabilistic modeling is an effective tool for evaluating team performance and predicting outcomes in sports. However, an important question that hasn’t been fully explored is whether these models can reliably reflect actual performance while assigning meaningful probabilities to rare results that differ greatly from expectations. In this study, we create an inference-based probabilistic framework built on expected goals (xG). This framework converts shot-level event data into season-level simulations of points, rankings, and outcome probabilities. Using the English Premier League 2015/16 season as a data, we demonstrate that the framework captures the overall structure of the league table. It correctly identifies the top-four contenders and relegation candidates while explaining a significant portion of the variance in final points and ranks. In a full-season evaluation, the model assigns a low probability to extreme outcomes, particularly Leicester City’s historic title win, which stands out as a statistical anomaly. We then look at the ex ante inferential and early-diagnostic role of xG by only using mid-season information. With first-half data, we simulate the rest of the season and show that teams with stronger mid-season xG profiles tend to earn more points in the second half, even after considering their current league position. In this mid-season assessment, Leicester City ranks among the top teams by xG and is given a small but noteworthy chance of winning the league. This suggests that their ultimate success was unlikely but not entirely detached from their actual performance. Our analysis indicates that expected goals models work best as probabilistic baselines for analysis and early-warning diagnostics, rather than as certain predictors of rare season outcomes.

\end{minipage}
\end{center}
}

% \keywords{Expected Goals, Football, Poisson Models, Sports Analytics, Win Probability}
% ---- Custom wide keywords block matching abstract ----
% Satisfy the class but print nothing

\maketitle
\thispagestyle{firstpagefooter}
\pagestyle{mystyle}      
% ---- Disable watermark after page 1 ----
\SetWatermarkText{}   % empty = removes text
\vspace{-1.0\baselineskip}
% ---- PREPRINT NOTE AFTER MAKETITLE ----
\begin{center}
\begin{tabular}{|c|}
\hline
\small \textbf{\textcolor{purple}{THIS IS A PREPRINT THAT HAS NOT YET UNDERGONE PEER REVIEW.}} \\
\hline
\end{tabular}
\end{center}

\section{Introduction}\label{sec1}

Uncertainty is a key part of all sports and plays a big role in attracting fans and keeping them engaged. To better understand and reduce this uncertainty, football has turned to data-driven methods. Football analytics marks a major change in how we interpret, evaluate, and improve the game. This drives the need for new measurements, like expected goals (xG) (Vilela 2024, Nipoti, and Schiavon 2025, Bandara et al. 2024). Beyond performance metrics, football analytics covers a wide range of tasks that can be tackled using statistical and predictive modeling. For example, it can predict match outcomes, assess player performance, examine team strategies, and guide decisions in player recruitment and injury prevention (see e.g., Souza et al. 2021, Skripnikov et al. 2025, Elsharkawi et al. 2025, title \& Suguna 2023). Together, these applications show the broad reach and growing importance of analytics in today’s game.

Expected goals (xG) models evaluate shot quality by calculating the chance of scoring based on past attempts (Spearman, 2018). In simple terms, expected goals assign a probability between 0 and 1 to every shot a team takes during a match. A score of 0 means that there is no chance that the shot being a goal, while 1 means that a goal is certain.

Formally, the expected goals (xG) of a team in a given match can be represented as the sum of the scoring probabilities of all its shots:  
\begin{equation}
\mathrm{xG}_{\text{team, match}} = \sum_{i=1}^{N} p_i,
\end{equation}

where $N$ shows the number of shots taken by the team in that match, and $p_i$ is the  probability that shot $i$ results in a goal.

This method is more effective than a conventional goal-based metric in addressing randomness in football, as a shot is a much more frequent occurrence than a goal (Anzer, \& Bauer, 2021). Historically, researchers have modeled the number of goals a team scores in a football match using statistical distributions to forecast match outcomes (Wheatcroft, 2021). For instance, goal-based approaches (Egidi \& Torelli 2021, Mead, O’Hare, and McMenemy) modelled the number of goals scored directly using statistical distributions such as Poisson-based models. The focus is on how many goals a team or a player is likely to score, while result based approaches directly model match outcomes (for example, win, draw or loss) rather than the sequence of events that leads to them (see, e.g., Macrì Demartino et al. 2024).
Since its development, the xG metric has become ubiquitous in the world of football. 

The majority of top-tier football teams and betting corporations employ these statistics, including related concepts of expected assists and post-shot expected goals. Additionally, these metrics play a critical role in player development and acquisition for organizations and in enhancing predictive models used in sports betting (Mead et al., 2023).
The primary purpose of these metrics is to provide a more comprehensive assessment of a player and team's performance beyond just the total number of goals scored. By quantifying shot probabilities, a team can gain an improved understanding of whether they are generating high-quality opportunities, experiencing poor finishing luck, or benefiting from favorable variance. This analytical tool has recently gained significant popularity, as the final outcome of a match does not always accurately reflect the opportunities that a team had.

Expected goals (xG) provide a more detailed picture of a team's performance in individual matches, but their broader relevance becomes clear when examined over an entire season. Discrepancies between cumulative xG and final league rankings highlight the concept of ranking uncertainty, whereby a team's position in the table may not accurately reflect its underlying performance. Even teams that consistently generate high-quality chances may underperform due to defensive errors, adverse variance, or unfavorable match dynamics. Conversely, other teams may outperform their xG by converting low-probability chances at unusually high rates. Such discrepancies expose structural limitations in point-based league standings as representations of team quality.

However, xG has limitations despite its widespread use and analytical utility. Methodological differences in data collection  approaches can yield substantially different results for identical shots when different xG models are employed. This variability requires a thorough examination of data sources and a clear understanding of each model’s underlying assumptions.

Another important limitation is that the predictive power of xG values in individual matches is limited, and subject to high variance. Match outcomes can deviate substantially from xG expectations due to randomness and the small sample sizes inherent in single games. Consequently, distinguishing real performance trends from statistical noise typically requires aggregating data across multiple matches to obtain meaningful xG-based insights. These limitations highlight the importance of interpreting xG metrics within a broader framework of uncertainty modeling and probabilistic reasoning in football analysis.

Nonetheless, much of the existing literature focuses primarily on point estimates of team or player performance, attention to the uncertainty and variability embedded in season-level outcomes derived from xG data. In particular, the disparity between cumulative xG performance and actual league standings remains a largely unexplored area, despite its relevance for accessing performance consistency, model reliability, and result fairness. This study addresses this gap by developing a probabilistic framework to quantify ranking uncertainty and identify rare match outcomes using metrics derived from xG. In other words, this study does not interpret xG as a deterministic predictor of final league outcomes. Instead, we distinguish between retrospective model fit and ex-ante inferential and diagnostic value of expected goals as an early-warning signal. Our central question is not whether xG can precisely anticipate rare outcomes, but whether it provides timely indications that a season may be unfolding in an unexpected direction.

Our primary objective is to conduct a systematic evaluation of how often teams deviate from their xG expectations in the final league table. We refer to this phenomenon as ranking uncertainty, where a team’s final position is inconsistent with its underlying xG-derived performance indicators. To assess this, we simulate alternative league standings by using xG-derived scoring intensities and compare them with the actual rankings, thereby identifying which teams overperform or underperform relative to expectations. 

Furthermore, we propose a framework for identifying and characterizing rare outcomes, such as underdog victories despite low expected goals (xG) or draws featuring highly imbalanced xG distributions. These anomalies challenge conventional modeling assumptions and call for careful analysis from both probabilistic and statistical perspectives.

Moreover, while several core predictors in football analytics (e.g., shot location and distance) are well established, this study introduces a set of hand-crafted, domain-informed features designed to enhance both model interpretability and tactical relevance. These features include spatial zones, body part used, shot placement, and match context, which are incorporated into a logistic regression framework to estimate the probability that a shot results in a goal. The resulting xG values are subsequently employed in Poisson-based scoring models to simulate match outcomes, which are then aggregated into full-season simulations to examine the variability and fairness of league rankings.

This study contributes a transparent and statistically grounded analytical pipeline that:
\begin{itemize}
\renewcommand\labelitemi{$\bullet$}
\item Quantifies ranking uncertainty across entire seasons through simulation-driven league tables based on expected goals (xG).
\item Provides empirical insight into the statistical signatures of rare or inconsistent match outcomes.
\item Examines the trade-off between predictive performance and data practicality when modeling imbalanced events.
\item Utilizes detailed shot-level and event-level data from the English Premier League (EPL) 2015/16, presenting a practical, real-world validation of the framework.
\end{itemize}
To the best of our knowledge, this is among the first studies to formally quantify league ranking uncertainty using xG simulations, while simultaneously addressing the modeling of rare football events within a competitive league context.

In light of these considerations, the remainder of this study is organized as follows: Section 2 reviews related literature on expected goals (xG) modeling and ranking evaluation. Section 3, introduces the proposed framework for constructing a logistic regression-based expected goals model using hand-crafted shot-level features derived from match event data. It also describes the estimation of ranking distributions and title probabilities across the 2015/16 EPL season, based on an xG-driven Poisson process. Section 4 presents a comparative analysis of predicted and actual league standings, including relevant performance metrics. Finally, Section 5 concludes with a discussion of the key findings, implications, and limitations of the study.
\section{Related Work} 
\subsection{Evolution of Expected Goals (xG)}	
The origins of expected goals (xG) models remain somewhat ambiguous, with most sources (e.g., Rathke 2017; Herbinet 2018; Umami et al. 2021) attribute the concept's foundation to Macdonald’s (2012) study of shot outcomes in ice hockey, while others (e.g., Spearman 2018) trace the term back to Green’s (2012) study. At its core, the expected goals framework can be formulated as a classification problem, as it involves estimating the probability that a shot results in a goal. Consequently, a range of statistical and  machine learning methods have been employed to model these probabilities, including multinomial logistic regression, gradient boosting, neural networks, support vector machines (SVM), and tree-based classification techniques (Anzer, \& Bauer, 2021).
\subsection{Features Used in xG Modeling}
Most features in these models are drawn from in-game data, which can generally be grouped into two categories: positional data and event-based data. Event-based data includes actions such as passes, duels, penalties, and shots, along with other on-field events recorded during a match. Each event is typically described by a set of variables for example, the pitch location (x and y coordinates), the area where the action was completed (such as the target of a pass or shot), the player involved, the match context, the outcome (success or failure), and a range of other situational details.

 Previous research has examined major football championships and national leagues to understand how goals are scored (Kubayi, 2020; Liu et al., 2015). In these studies, researchers analysed variables such as league ranking (top, mid, or lower-tier clubs), match outcome (win, draw, or loss), venue (home or away), and tactical system. In another study (Rathke, 2017), projected that presenting players with their xG values could improve attacking performance by highlighting the most effective shot locations and techniques. The study found that both shooting distance and angle predictors significantly influence the probability of scoring a goal. Similarly, Spearman (2018) examined the effects of distance and angle on shooting outcomes and emphasised that shot location is a critical feature to incorporate into predictive models. Given its significance in explaining the variability of goal probability, shot location remains a central variable in nearly all studies related to expected goals (e.g., Kharrat et al. 2020, Brechot \& Flepp. 2020).

Another commonly discussed feature in the literature is shot type, which provides contextual information about the attempt and can be divided into two subcategories. The first subcategory relates to the body part used for the shot (left or right foot, head, or another part). The second concerns the game situation in which the shot occurs. Depending on the model, this may include open play, counterattack, free kick, or penalty kick. Brechot and Flepp (2020) incorporated these features into their xG model and found that both factors significantly affect shot outcomes. Specifically, they reported that shots taken from a free kick are more likely to result in goals, penalties even more so, whereas headers are considerably less likely to do so compared with shots taken in open play with either foot.

Beyond these technical and spatial features, recent studies have emphasized the importance of incorporating dynamic, match-level context into xG modeling. One such factor, match status, has been identified as a critical variable in explaining possession patterns throughout the match, as teams' strategies evolve with the scoreline. For instance, a losing-match status is often associated with increased ball possession as teams seek to regain control and build sustained attacks through indirect play. Conversely, a winning-match status typically corresponds to reduced possession, reflecting a strategic preference for direct play or counterattacking (Wang et al., (2022)). 

Furthermore, match status also affects possession by field zone. Lago (2009) found that possession time in the offensive third was longer when a team was losing than when it was winning or drawing. Nevertheless, due to factors such as limited sample size, inconsistent analytical frameworks, and the inherent complexity of football, as a dynamic and unpredictable sport, much of the existing literature reports inconclusive findings regarding these relationships..

\subsection{League-Level xG Modeling and Simulation}
The English Premier League (EPL) is widely regarded as one of the most competitive and financially valuable football leagues in the world (Cox and Philippou 2022). The 2015/16 EPL season, however, was highly atypical: Leicester City, a club that had narrowly avoided relegation the previous year and began the campaign with bookmakers’ odds of 5000-1, remarkably secured the league title. Several factors contributed to this outcome, including an exceptionally efficient conversion rate, a robust defensive organization, and consistent performances against higher-resourced opponents. This unprecedented achievement not only captured global attention but also challenged conventional assumptions about the determinants of success in football. In particular, Leicester’s triumph underscores the limitations of conventional point-based standings in accurately reflecting team performance. 

Recent studies have introduced Poisson-based modeling frameworks for simulating match outcomes using xG inputs or goal expectations (e.g., Eggels et al., 2016; Nguyen, 2021). However, these works have mainly focused on predicting individual match results or final point totals, without examining their relationship to full league standings or identifying systematic over- and underperformance across an entire season.

This study addresses this gap by posing the following research question: How accurately can expected goals (xG)-based models simulate full-season league standings, and what insights do they provide into ranking uncertainty and rare football outcomes in the English Premier League.

\section{Material and Methods}
This study addresses this research objectives by developing a statistical framework that leverages xG-based inference to simulate season-long performance, analyze ranking distributions, and compare predicted versus actual league tables. Unlike previous approaches that emphasize single-match forecasting, our study quantifies ranking uncertainty, identifies performance discrepancies, and evaluates the reliability and fairness of xG-driven assessments in a full-season competitive context. Figure 1 illustrates the architectural pipeline of proposed xG-based framework for season-level ranking analysis.

\begin{figure}[H]
  \centering
  \makebox[\linewidth][c]{%
    \includegraphics[width=1.4\linewidth]{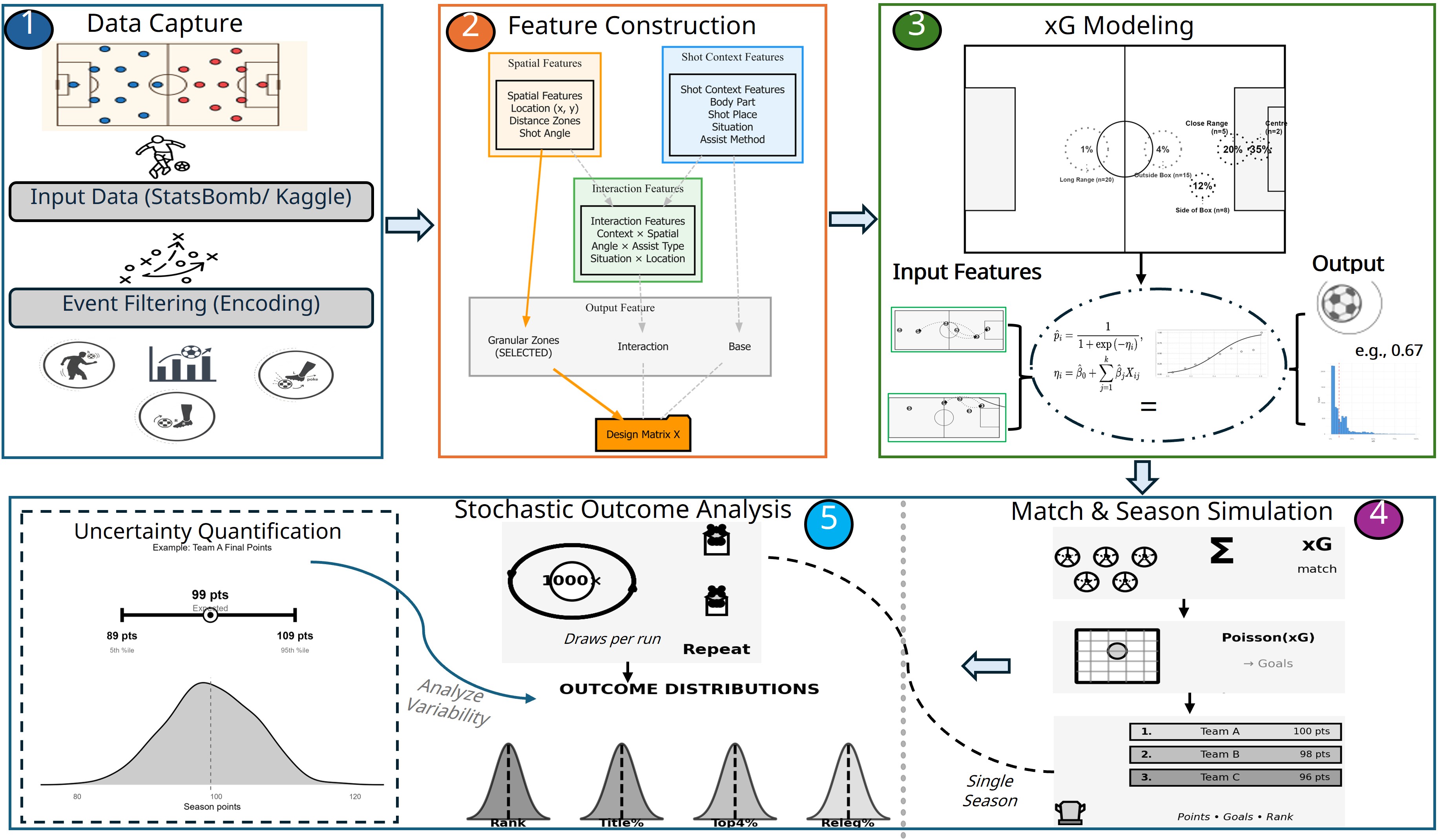}}
  \vspace{0.7cm}
  \caption{End-to-end xG-based framework for season-level ranking simulation and inference}
  \label{fig:framework}
\end{figure}

\subsection{Data Source}
The research utilizes the English Premier League (EPL) 2015/16 football dataset (Secărean, 2024). The data comprises two main components:
\begin{itemize}
\renewcommand\labelitemi{$\bullet$}
\item Match-level data: Each observation represents a single match and includes attributes such as home and away team names, final scores, match date, and betting odds.

\item Event-level data: This dataset contains detailed, timestamped in-game events such as shots, fouls, and passes. Each event is linked to a match via a unique match ID (id\_odsp). In this study, we used a subset of variables relevant to modeling and simulation summarized in Table 1. Variables not included in the analysis (e.g., secondary event types, redundant timestamps) were excluded to reduce data complexity and improve model efficiency.

\end{itemize}

\renewcommand{\tablename}{Table}
\begin{table}[h!]
\centering
\caption{Description and type of the variables considered in this study.}
\begin{tabular}{lll}
\toprule
\textbf{Variable} & \textbf{Description} & \textbf{Type} \\
\midrule
id\_odsp & Unique match ID used to join match and event data & Categorical (ID) \\
event\_team & Team taking the shot & Categorical \\
side & Home (1) or Away (2) indicator & Binary \\
is\_goal & Indicator if the shot resulted in a goal & Binary \\
location & Coded pitch zone where the shot occurred & Categorical \\
shot\_place & Shot placement (e.g., bottom left, top right) & Categorical \\
bodypart & Body part used for the shot (e.g., left foot, head) & Categorical \\
situation & Match context: open play, corner, free kick & Categorical \\
assist\_method & Type of assist (pass, cross, etc.) & Categorical \\
fast\_break & Whether the shot came from a counterattack & Binary \\
ht, at & Home and away team names & Categorical \\
fthg, ftag & Full-time home and away goals & Numeric \\
\bottomrule
\end{tabular}

\end{table}
Furthermore, this study analyzes all 380 matches from the EPL 2015/16 season, comprising 20 teams competing over 38 rounds. Event-level data were collected for both teams in each match, resulting in 760 team-level observations. Across a season, a total of 975 goals were recorded, including 172 home wins, 93 draws, and 115 away wins. We use this data to build the basis for full-season simulations, in which Poisson-distributed xG values are employed to evaluate ranking uncertainty and performance deviations.

\subsection{Feature Engineering}
To estimate the probability that a shot results in a goal (i.e., expected goals or xG), several shot-level features were extracted and transformed from the event dataset. The data were first filtered to include only shot events (\texttt{event\_type == 1}) and complete cases for all relevant predictors. The following categorical predictors were engineered and incorporated into the logistic regression model:

\begin{itemize}
\renewcommand\labelitemi{$\bullet$}
\item \textit {location}: Pitch areas were grouped into distance-based zones, such as \textit {close range}, \textit {medium range}, \textit {Outside box}, and \textit {long range}. A more granular version was also tested, distinguishing \textit {centre of the box} from \textit {side of the box}.
\item \textit {shot\_place}: The intended direction of the shot (e.g., top right corner, bottom left) was encoded as a categorical variable to capture accuracy and targeting behavior.
\item \textit {bodypart}: The body part used to take the shot—right foot, left foot, or head—was included due to its known effect on scoring likelihood.
\item \textit {situation}: The match context in which the shot occurred, such as open play, set piece, corner, or free kick.
\item \textit {assist\_method}: The type of assist preceding the shot, including pass, cross, through ball, or header.
\item \textit {fast\_break}: A binary indicator denoting whether the shot followed a counterattack.
\end{itemize}
Additionally, an interaction term was introduced as a new categorical feature, distance\_zone, derived from the \textit {location} variable to represent spatial shooting zones. This feature groups shot locations into distance-based categories to enhance interpretability and capture spatial effects on goal probability. To further explore potential dependencies between spatial and biomechanical variables, we computed interaction terms (e.g., \texttt{distance\_zone * bodypart}). All categorical variables were appropriately factorized and included directly in the model, \textit{glm()} function internally handles categorical encoding.

These features were selected based on prior empirical findings and their domain relevance, and were used as inputs in a logistic regression model for estimating xG at the shot level. Table 2 outlines the feature engineering steps used to define the distance zone predictor.

\renewcommand{\tablename}{Table}
\begin{table}[h!]
\centering
\caption{Distance Zone Feature Engineering Steps}
\begin{tabularx}{\textwidth}{clX}
\toprule
\textbf{Step} & \textbf{Task} & \textbf{Details} \\
\midrule
1 & Input Data & Start with filtered shot events (\texttt{event\_type == 1}). \\
2 & Identify Location & Use the location variable from event data (coded 1--19). \\
3 & Define Zones & Group location codes into categories based on proximity to goal. \\
4 & Zone Mapping & Close Range: 10, 12, 13, 14; \newline
                   Medium Range: 3, 9, 11; \newline
                   Outside Box: 15, 16; \newline
                   Long Range: 17, 18; \newline
                   Other: all remaining codes. \\
5 & Create New Variable & Assign new variable \texttt{distance\_zone} based on the above mapping. \\
\bottomrule
\end{tabularx}
\end{table}

\subsection{Model Development and xG Estimation}
The central objective of this modeling step was to estimate the probability that a shot results in a goal, commonly referred to as expected goals (xG). We formulated this as a binary classification problem, where each shot is labeled either as a goal ($y_i = 1$) or not ($y_i = 0$). 

To model this probability, we applied logistic regression, a generalized linear model with a binomial distribution and logit link function. The logistic regression estimates the probability $\hat{p}_i$ that a shot $i$ results in a goal, based on a linear combination of categorical predictors. The model can be expressed as shown in Eq. (2):

\begin{equation}
\text{logit}(p) = \log\!\left(\frac{p}{1 - p}\right) 
= \beta_{0} + \sum_{j=1}^{k} \beta_{j} X_{j}.
\end{equation}

\noindent
Where: \\
\quad $p$ : probability that a shot is converted into a goal (expected goal, xG) \\  
\quad $X_{j}$ : $j$-th predictor variable \\  
\quad $\beta_{j}$ : coefficient corresponding to the $j$-th predictor \\  
\quad $\beta_{0}$ : intercept term  
\vspace{1\baselineskip} 

\noindent Table 3 presents the definitions of the outcome, covariates, and model parameters used in the analysis.
\vspace{-\baselineskip} % removes one line of space
\renewcommand{\tablename}{Table}
\begin{table}[h!]
\centering
\caption{Definitions of outcome, covariates, and model parameters.}
\begin{tabularx}{\textwidth}{lX}
\toprule
\textbf{Symbol} & \textbf{Definition} \\
\midrule
$y_i$ & Observed outcome of shot $i$; $1 =$ goal, $0 =$ no goal \\
$p_i$ & Probability of a goal (expected goals, xG) for shot $i$ \\
$\text{logit}(p_i)$ & Log-odds transformation of the probability \\
$\eta_i$ & Linear predictor (sum of weighted predictors) for shot $i$ \\
$\beta_{0}$ & Intercept term in the regression model \\
$\beta_{j}$ & Coefficient for the $j$-th feature (e.g., body part, location) \\
$X_{ij}$ & Value of the $j$-th feature for shot $i$ \\
\bottomrule
\end{tabularx}
\end{table}
\vspace{-\baselineskip} % removes one line of space
\subsubsection{Model 1: Base Model with Distance Zones}
The base specification models the log-odds of a shot resulting in a goal as a linear function of categorical covariates that describe the shot context. The covariates include distance zone (close, medium, outside box, long range), an engineered feature that groups pitch locations by approximate shooting distance (see Eq. (3)). Additional predictors include:

\renewcommand\labelitemi{$\bullet$}
\begin{itemize}
    \item \textit{shot\_place}: (e.g., bottom left, top right), capturing the intended target area.
    \item \textit{bodypart}: (right foot, left foot, head), representing the biomechanics of finishing.
    \item \textit{situation}: (open play, corner, free kick, penalty), reflecting match context.
    \item \textit{assist\_method}: (pass, cross, rebound, etc.), describing chance creation.
    \item \textit{fast\_break}: (yes/no), indicating counterattack situations.
    \vspace{-\baselineskip}
\end{itemize}
\begin{align}
\text{logit}(p) = \beta_{0} 
&+ \beta_{1} \cdot \text{distance\_zone} 
+ \beta_{2} \cdot \text{shot\_place} 
+ \beta_{3} \cdot \text{bodypart} \notag \\
&+ \beta_{4} \cdot \text{situation} 
+ \beta_{5} \cdot \text{assist\_method} 
+ \beta_{6} \cdot \text{fast\_break}.
\end{align}
\vspace{-\baselineskip}
\subsubsection{Model 2: Interaction Model (Distance Zone × Body Part)}
To capture biomechanical interactions between spatial context and finishing technique, Model 2 extends the base specification by including an interaction between distance zone and body parts. This interaction allows the model to reflect, for example, that headers are more effective from close range but rarely successful from long distances. The model is expressed as:
\begin{align}
\text{logit}(p) = \beta_{0} 
&+ \beta_{1} \cdot \text{distance\_zone} 
+ \beta_{2} \cdot \text{shot\_place} 
+ \beta_{3} \cdot \text{bodypart} \notag \\
&+ \beta_{4} \cdot \text{situation} 
+ \beta_{5} \cdot \text{assist\_method} 
+ \beta_{6} \cdot \text{fast\_break} 
+ \beta_{7} \cdot \bigl(\text{distance\_zone} \times \text{bodypart}\bigr).
\end{align}
\subsubsection{Model 3: Granular Spatial Zones}
The third model increases spatial resolution by replacing the coarse distance zones with a more detailed categorical variable, {distance\_zone\_granular} (see Eq. (5)). This feature distinguishes finer pitch locations such as the \textit {centre of the box}, \textit {side of the box}, \textit {penalty spot}, and \textit {very close range}. The objective is to assess whether greater spatial granularity improves predictive performance relative to simpler distance groupings.

\begin{align}
\text{logit}(p) = \beta_{0} 
&+ \beta_{1} \cdot \text{distance\_zone\_granular} 
+ \beta_{2} \cdot \text{shot\_place} 
+ \beta_{3} \cdot \text{bodypart} \notag \\
&+ \beta_{4} \cdot \text{situation} 
+ \beta_{5} \cdot \text{assist\_method} 
+ \beta_{6} \cdot \text{fast\_break}.
\end{align}
Model performance was compared across all three specifications using Akaike Information Criterion (AIC), residual deviance, and the number of predictors used (see Table 4).
\vspace{-\baselineskip}

\renewcommand{\tablename}{Table}
\begin{table}[h!]
\centering
\caption{Model comparison by AIC, residual deviance, and number of predictors.}
\label{tab:model_comparison}
\begin{tabularx}{\textwidth}{lccc}
\toprule
\textbf{Model} & \textbf{AIC} & \textbf{Residual Deviance} & \textbf{Num. Predictors} \\
\midrule
Base (distance\_zone) & 5354.47 & 5334.47 & 10 \\
With Interaction       & 5357.31 & 5329.31 & 14 \\
Granular Zones         & 5261.82 & 5239.82 & 11 \\
\bottomrule
\end{tabularx}
\end{table}
Among the three xG models, the \textit{Granular Spatial Zones} model achieved the lowest AIC and residual deviance ($\Delta$AIC $\geq 90$), indicating the best parsimony-adjusted fit.

\subsubsection{Model Prediction and xG Assignment}
The logistic regression model was trained using the selected optimal covariates, and the parameters were estimated using the maximum likelihood estimation (MLE). The fitted model was subsequently applied to compute the xG value for each shot, defined as the model-predicted probability that a given attempt results in a goal, conditional on its characteristics.

The fitted logistic regression model produces a predicted probability $\hat{p}_i$ for each shot $i$, given its input features (see Eqs. (6)–(7)), computed using the inverse logit (sigmoid) function applied to the linear predictor $\eta_i$:
\begin{equation}
\hat{p}_i = \frac{1}{1 + \exp(-\eta_i)},
\end{equation}
\begin{equation}
\eta_i = \hat{\beta}_{0} + \sum_{j=1}^{k} \hat{\beta}_{j} X_{ij}.
\end{equation}
Where:
\begin{description}
  \item[$\hat{p}_i$:] estimated probability of a goal (i.e., xG value).
  \item[$\eta_i$:] linear predictor combining model coefficients and input features.
  \item[$\hat{\beta}_{0}$:] model intercept.
  \item[$\hat{\beta}_{j}$:] estimated coefficient for predictor $X_j$.
  \item[$X_{ij}$:] value of the $j$-th feature for shot $i$.
\end{description}

The simulation framework is summarized in Table~\ref{tab:algorithm_steps}.

\begin{table}[!htbp]
\centering
\caption{Ex ante simulation framework for season outcome inference}
\label{tab:algorithm_steps}
\begin{tabularx}{\textwidth}{l l X}
\toprule
\textbf{Stage} & \textbf{Component} & \textbf{Specification} \\
\midrule
\multirow{4}{*}{Preparation}
 & Temporal split 
 & Split into first half  and second half fixtures. \\
 
 & xG estimation 
 & Shot-level logistic model $\rightarrow$ match-level xG for first-half matches only. \\
 
 & Team strength 
 & Aggregate first-half xG to team-level attack ($xG^{\text{for}}$) and defense ($xG^{\text{against}}$). \\
 
 & Normalization 
 & Convert xG totals to per-match rates; normalize by league means. \\
\midrule
\multirow{3}{*}{Simulation}
 & Poisson intensities 
 & Opponent-adjusted scoring rates $\lambda_{\text{home}}, \lambda_{\text{away}}$ based on attack and defense. \\
 
 & Goal generation 
 & $G_{m}^{h} \sim \text{Poisson}(\lambda_{m}^{h}),\;
    G_{m}^{a} \sim \text{Poisson}(\lambda_{m}^{a})$. \\
 
 & Standings update 
 & Assign points and goals; combine simulated second half with realised first half. \\
\midrule
\multirow{2}{*}{Inference}
 & Ranking 
 & Rank teams per simulation using league tiebreakers. \\
 
 & Outcome probabilities 
 & Compute $\mathbb{E}[\text{Points}]$, $\mathbb{E}[\text{Rank}]$, 
   $P(\text{Title})$, $P(\text{Top4})$, $P(\text{Releg})$. \\
\bottomrule
\end{tabularx}
\end{table}

\section{Results}
This section evaluates the accuracy of the xG-based model in simulating team performance and capturing ranking uncertainty during the 2015/16 EPL season. We present statistical comparisons, visualization of predicted vs actual values, and uncertainty quantification across 1000 simulations.

\subsection{Mid-season xG as an Early-warning Signal}
To assess whether expected goals provide early diagnostic information, we analyze team performance at the mid-season point of the campaign. Teams are grouped according to cumulative first-half xG, and their subsequent second-half performance is examined.

Figure 2 shows a clear monotonic relationship between first-half xG quartiles and second-half point accumulation, indicating that xG contains systematic information about subsequent performance.
% --- Figure 2 inline (no floating) ---
\renewcommand{\figurename}{Figure}
\begin{center}
\includegraphics[width=\linewidth]{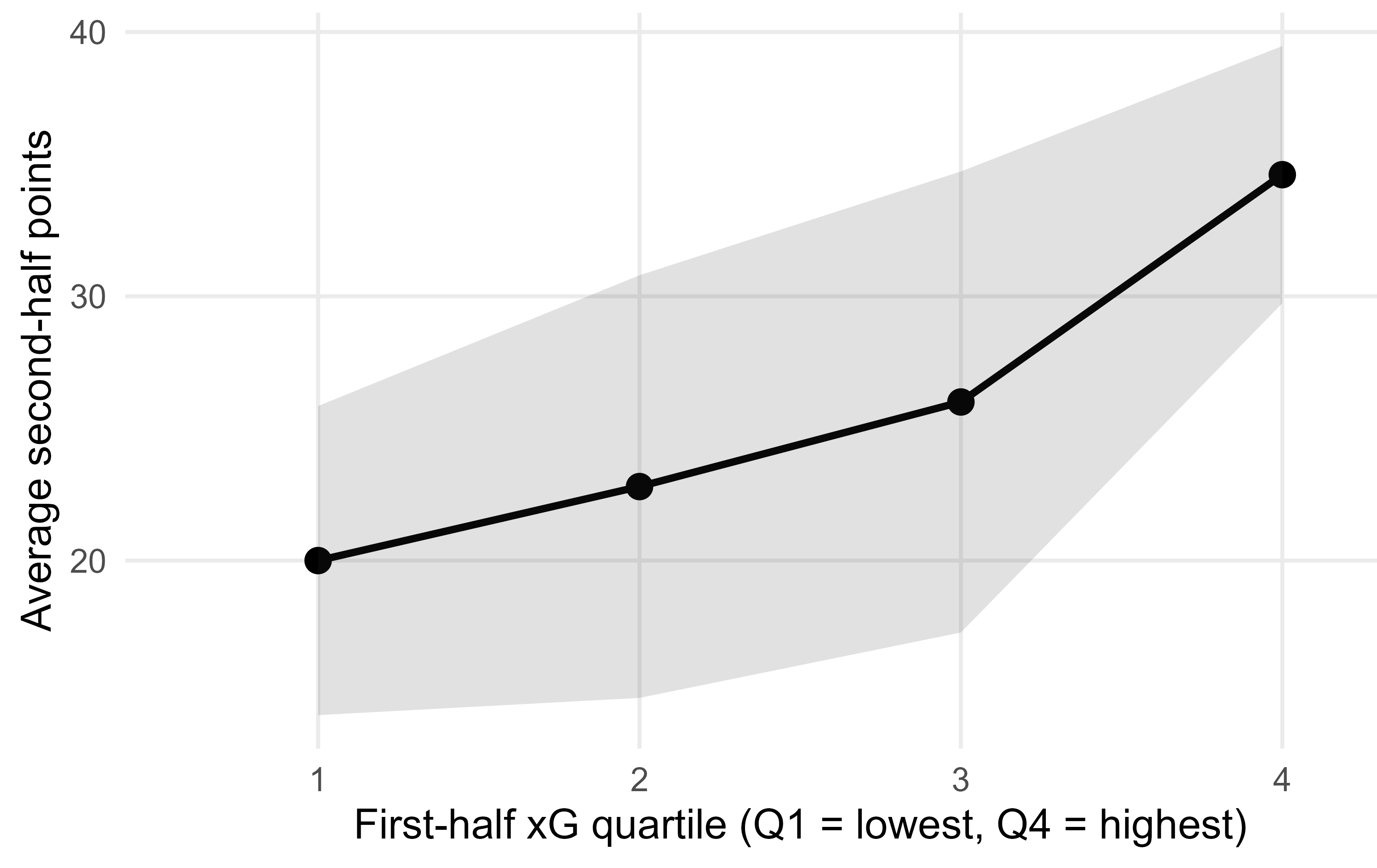} 
\captionof{figure}{Teams grouped into quartiles by mid-season xG.}
\label{fig:pred_vs_real}
\end{center}
Importantly, this relationship persists even after controlling for mid-season points (see Figure 3), suggesting that xG captures dimensions of underlying performance not fully reflected in the league table at that stage.

% --- Figure 3 inline (no floating) ---
\renewcommand{\figurename}{Figure}
\begin{center}
\includegraphics[width=\linewidth, height=0.35\textheight, keepaspectratio]{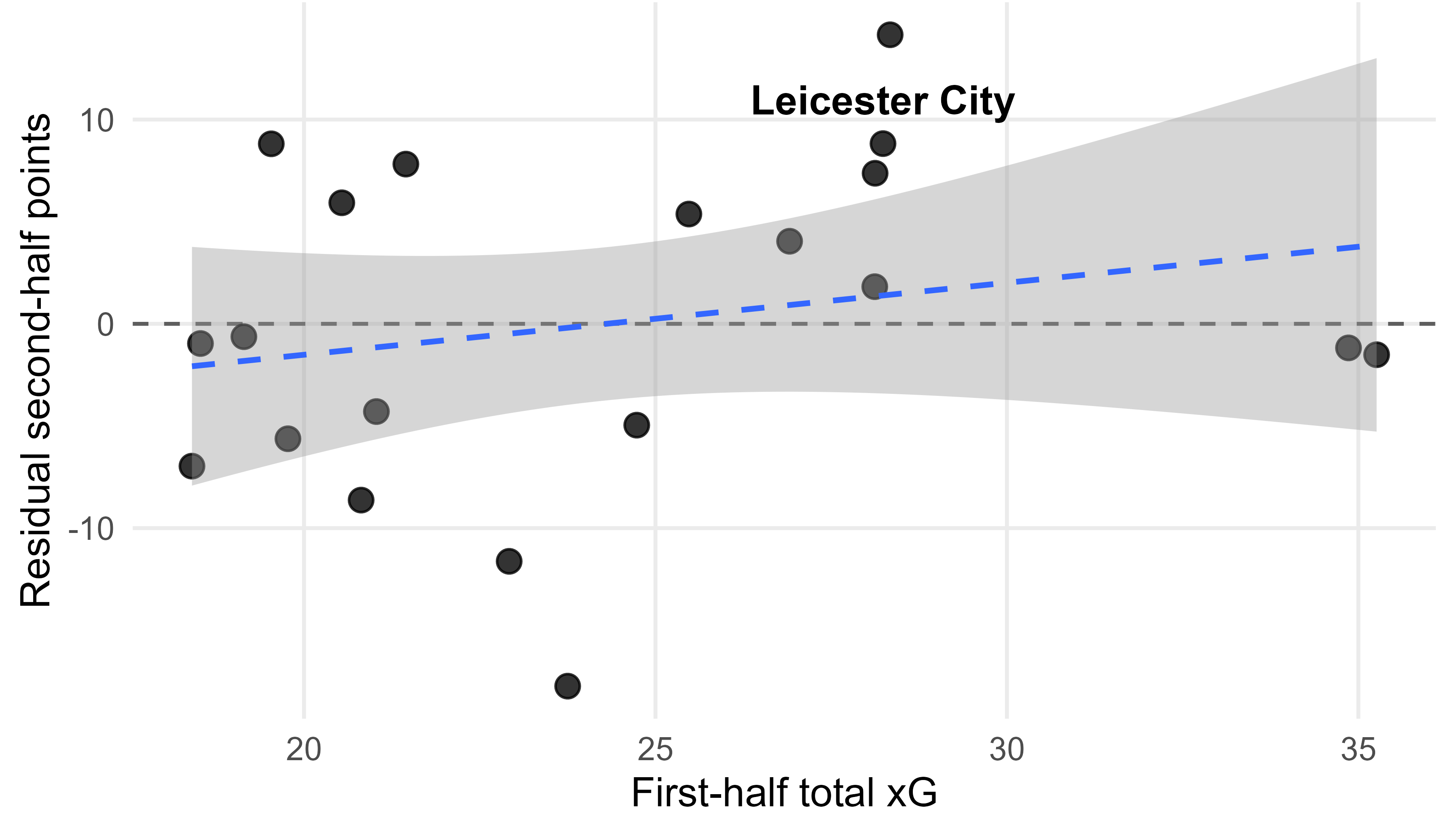} 
\captionof{figure}{Residualised second-half points vs first-half xG.}
\label{fig:pred_vs_real}
\end{center}
At mid-season, Leicester’s league position was 2nd and Arsenal led the table on rank. In contrast, Leicester City ranked fourth in xG at mid-season, despite leading the league on points (see figure 4).
% --- Figure 4 inline (no floating) ---
\renewcommand{\figurename}{Figure}
\begin{center}
\includegraphics[width=\linewidth, height=0.5\textheight, keepaspectratio]{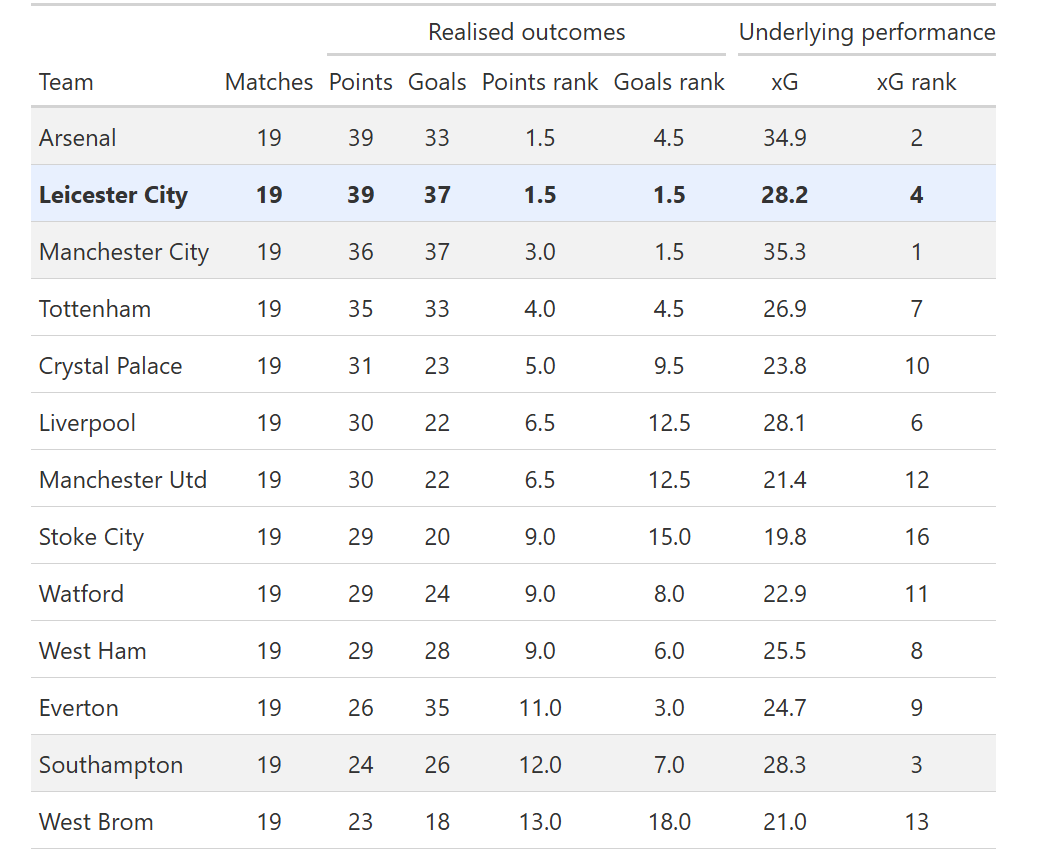}
\captionof{figure}{Mid-season rank table realised outcomes vs underlying performance}
\label{fig:pred_vs_real}
\end{center}
The mid-season  shows the moderate discrepancy between points-based and xG-based rankings and reveals that Leicester’s subsequent success was not entirely detached from their underlying performance, even though their eventual championship remained a highly improbable outcome under the model (see Appendix~C).

\subsection{Mid-Season xG and Ex Ante Outcome Probabilities}
To further assess the ex ante inferential value of expected goals, we simulate the remainder of the season using only information available at the mid-season point. Specifically, team attacking strength is derived from cumulative first-half xG, which is converted into a per-match scoring rate and used to simulate second-half match outcomes. First-half realised points are carried forward, while all second-half results are generated probabilistically.

\subsubsection{Mid-Season Diagnostic Signal: xG versus Realised Performance}

We begin by examining the relationship between expected goals and realised league outcomes at mid-season. Figure~5 compares team rankings based on cumulative first-half xG with rankings based on realised points after 19 matches. Teams close to the diagonal exhibit consistency between underlying performance and results, whereas deviations indicate over- or underperformance relative to xG.

Several teams occupying top positions in the points table—such as Arsenal and Manchester City also rank highly in xG, suggesting that their league positions were supported by strong underlying performance. Leicester City, who were third in the league at mid-season, likewise ranked highly in cumulative xG, indicating that their early success was not purely anomalous. By contrast, some teams accumulated points disproportionate to their xG, signalling potential regression risk in the second half of the season.
% --- Figure 5 inline (no floating) ---
\renewcommand{\figurename}{Figure}
\begin{center}
\includegraphics[width=\linewidth, height=0.5\textheight, keepaspectratio]{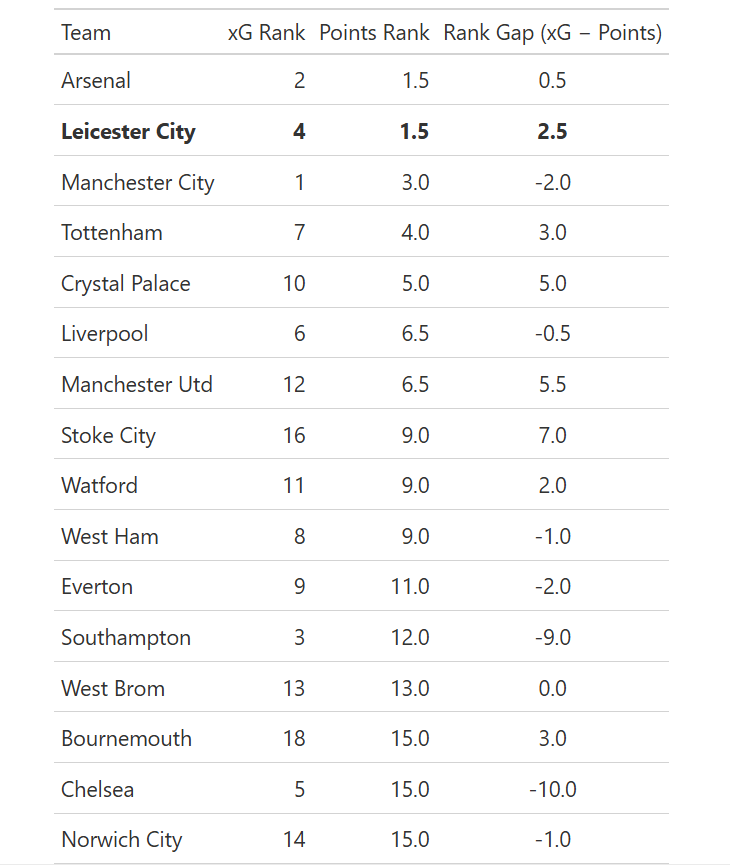}
\captionof{figure}{Mid-season rank-gap diagnostic }
\label{fig:pred_vs_real}
\end{center}

\subsubsection{Predicted Second-Half Points with Uncertainty}

Using first-half xG only, we simulate the remaining half fixtures of the season 1,000 times under a Poisson-based parametrized process calibrated on team-level attacking and defensive strength. Figure~6 presents the predicted second-half points for each team, alongside 95\% confidence intervals derived from the simulation distribution.

The results indicate substantial heterogeneity in uncertainty across the league. Teams with strong and stable xG profiles such as Arsenal, Manchester City, and Tottenham exhibit relatively narrow confidence intervals, reflecting consistent underlying performance. In contrast, mid-table teams display wider intervals, indicating greater sensitivity to stochastic match outcomes.

 In other words, most teams' realised outcomes fall within the central mass of their predicted distributions, indicating that mid-season xG captures underlying performance strength.

% --- Figure 6 inline (no floating) ---
\begin{center}
\includegraphics[width=\linewidth]{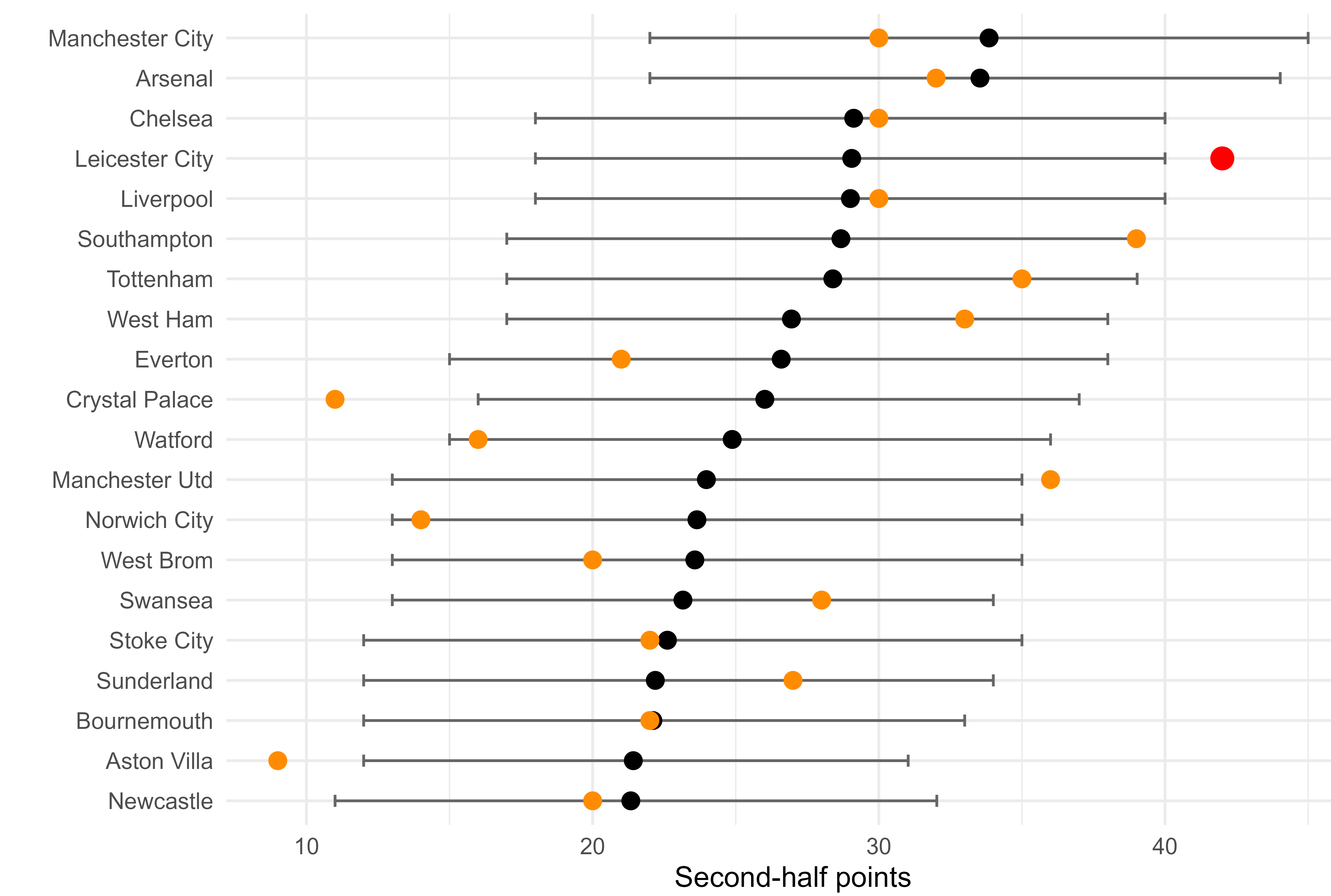}
\vspace{0.8em}
\captionof{figure}{Ex ante simulations conditional on mid-season xG, (where black is the mean simulated points, grey is 95\% and orange is the realised outcome)}
\label{fig:pred_vs_real}
\end{center}
Leicester City’s predicted second-half points lie above the league median, but well below their eventual realised performance. Importantly, their realised second-half points fall within the upper tail of the simulated distribution rather than entirely outside it, indicating that their title-winning run represents a low-probability but feasible outcome under mid-season xG information.
\subsubsection{Prediction Error and Agreement at Mid-Season}

To assess agreement between predicted and realised second-half outcomes, we examine the difference between simulated mean points and realised second-half points. Most teams cluster around zero, indicating that first-half xG provides a reasonable baseline for second-half performance on average. However, several notable deviations remain. Leicester City emerges as the most pronounced positive outlier, substantially exceeding their predicted mean. Conversely, teams such as Chelsea and Manchester United underperform relative to their xG-based expectations, suggesting inefficiencies or contextual factors not fully captured by chance quality alone.

These deviations underscore the probabilistic nature of the framework xG constrains the distribution of plausible outcomes but does not eliminate variance arising from finishing efficiency, tactical adaptation, injuries, or random match events.

\subsubsection{Mid-Season Rank Inference and Uncertainty}

Beyond point totals, the simulation framework enables inference over league rankings. Figure~7 compares predicted average finishing positions from the mid-season simulations with the realised final standings. 

The model successfully distinguishes broad performance tiers, separating title contenders, mid-table teams, and relegation candidates. Leicester City’s predicted average rank at mid-season lies within the top four, indicating that their underlying performance justified contender status even if the eventual championship outcome lay in the extreme tail of the distribution.

% --- Figure 7 inline (no floating) ---
\begin{center}
\includegraphics[width=\linewidth]{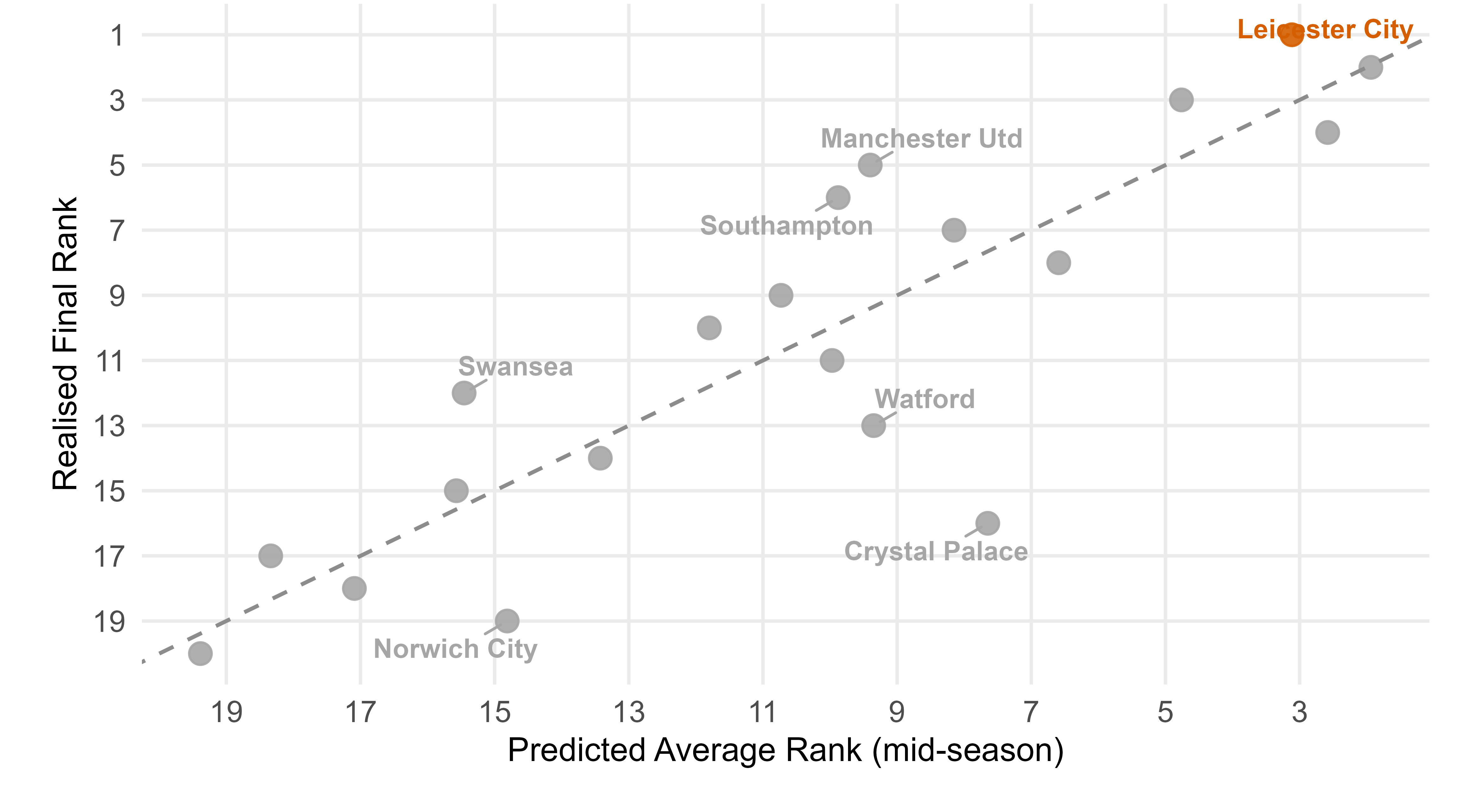}
\vspace{0.8em}
\captionof{figure}{Mid-Season Predicted Average Rank vs Realized Final Rank}
\label{fig:pred_vs_real}
\end{center}
In addition, the figure reveals that mid-season xG simulations correctly captured league tiers, even if they did not predict the exact final ordering.
On the other hand, rank variability across simulations is lowest for teams at the top and bottom of the table, and highest for mid-table teams, reflecting greater competitive balance and uncertainty in those regions of the league.

\subsubsection{Ex Ante Outcome Probabilities}

A central advantage of the simulation-based approach is the ability to estimate probabilistic outcomes rather than point forecasts. Figure~7 reports the mid-season probabilities of winning the title, finishing in the top four, and being relegated for all teams.
In addition, the figure illustrates how first-half expected-goals performance shapes simulated second-half outcomes. Several teams traditionally viewed as title contenders, such as Arsenal and Manchester City, retain the highest expected points and title probabilities. Leicester City also emerges as a statistically plausible contender at mid-season, with an average simulated rank of 3.1, a title probability of 16.7\%, and a top-four probability exceeding 80\%.

% --- Figure 5 inline (no floating) ---
\begin{center}
\includegraphics[width=\linewidth]{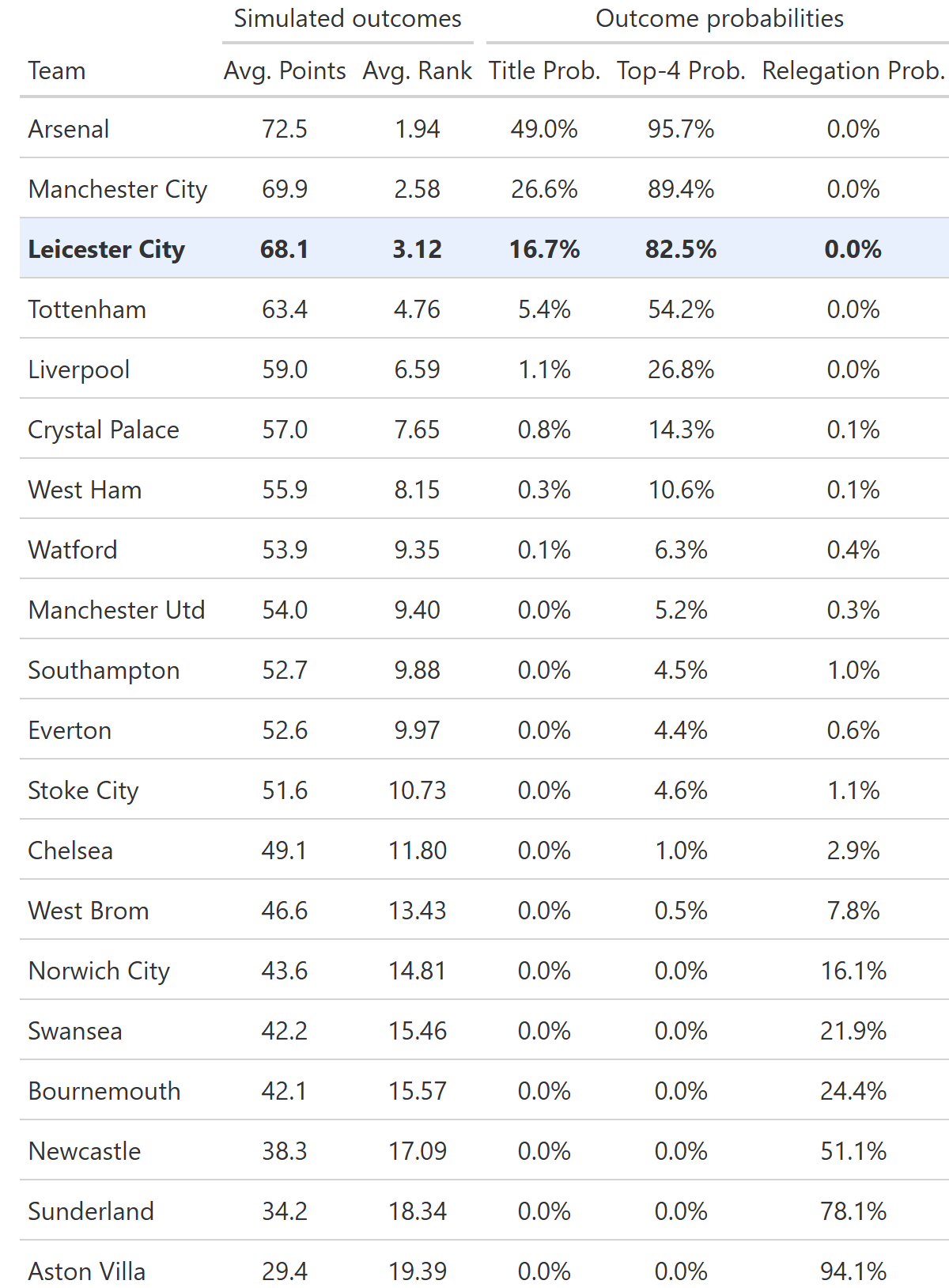}
\vspace{0.8em}
\captionof{figure}{Second-half simulations conditional on first-half expected goals}
\label{fig:pred_vs_real}
\end{center}

At the bottom of the table, teams such as Aston Villa, Sunderland, and Newcastle exhibit high relegation probabilities, aligning closely with their eventual outcomes.

\subsubsection{Leicester City: From Mid-Season Signal to Tail Outcome}

Leicester City provides a compelling illustration of the distinction between ex ante inference and realised outcomes. At mid-season, their strong xG performance placed them among the league’s top teams in terms of underlying quality. Simulations based on this information assigned them a meaningful probability of a top-four finish and a small but non-negligible chance of winning the title.

Their eventual championship, however, represents an extreme realisation within the upper tail of the predicted distribution.

Taken together, the mid-season results demonstrate that expected goals are more valuable as an early-warning and uncertainty-quantification tool. They provide a disciplined baseline for inference under partial information, while explicitly allowing for the rare but consequential events that define football seasons.

\subsection{Ex Post Benchmark: Full-Season xG Simulations}
For completeness, we evaluate the simulation framework retrospectively using full-season xG inputs. Because this benchmark conditions on information unavailable at mid-season, it represents a best-case upper bound on achievable model fit rather than a feasible real-time forecast.

Overall, the framework reproduces the broad structure of the league table, with strong rank correlations and moderate point-level errors (see table 6 in Appendix A). Simulated point totals typically fall within two to three match outcomes of realised season totals, while average ranking errors are approximately three league positions.

However, extreme outcomes most notably Leicester City’s championship—remain difficult to anticipate even under full-season information. Accordingly, this ex-post benchmark serves primarily as a reference point against which the mid-season ex-ante results should be interpreted, reinforcing the role of xG as a probabilistic baseline and early-warning diagnostic rather than a deterministic predictor.

\subsection{Ranking Uncertainty and Simulation Spread}
To quantify the variation in league rankings that may arise from the inherent randomness of match outcomes even when underlying xG performance remains consistent we analyzed the distribution and variance of team ranks across 1,000 full-season simulations. This ranking variability captures the uncertainty in seasonal outcomes and highlights which teams exhibited stable versus variable predicted performance. This step is critical for ensuring that team rank distributions converge as the number of simulations increases, rather than fluctuating erratically due to sample size limitations. The results demonstrate that:

\begin{itemize}[label=\textbullet]
    \item For most teams, the simulated rank values converge early, with only minimal shifts observed after ~500 simulations (see Appendix B).
    \item Top- and bottom-ranked teams, such as Arsenal, Manchester City, Aston Villa, and Norwich, exhibit particularly stable rank curves, indicating low variance in their simulated season outcomes.
    \item Mid-table teams show slightly greater variation in the early iterations but still stabilize by around 750–1,000 iterations, confirming that the sample size is sufficient for reliable predictions.

\end{itemize}

\subsection{Predicted Title, Top-4, and Relegation Probabilities (Mid-Season Ex Ante)}
A major strength of probabilistic simulation is its ability to quantify uncertainty in team outcomes under partial information. Beyond average rank or points, these simulations allow us to estimate the likelihood that a team achieves critical league milestones such as winning the title, qualifying for the top four, or being relegated. All results reported in this section are generated ex ante at mid-season, using expected-goals information from the first half of the season only.

Figure~9 presents the mid-season outcome probabilities of winning the title, finishing in the top four, and being relegated for all 20 EPL teams in the 2015/16 season. Arsenal entered the mid-season point with the highest title probability (49.0\%), followed by Manchester City (26.6\%) and Tottenham (5.4\%), confirming their status as strong xG-based contenders at that stage of the campaign.

In contrast to narratives based solely on league position, Leicester City also emerges as a statistically plausible contender at mid-season. The model assigns Leicester an average simulated rank of 3.12, a title probability of 16.7\%, and a top-four probability exceeding 80\%. While these values are lower than those of Arsenal and Manchester City, they indicate that Leicester’s underlying performance supported contender status, rather than representing a purely anomalous overperformance.

At the lower end of the table, teams such as Aston Villa, Sunderland, and Newcastle United exhibit the highest predicted probabilities of relegation, reflecting persistently weak xG profiles at mid-season. In particular, Aston Villa’s relegation probability exceeds 90\%, while both Sunderland and Newcastle face probabilities above 50\%, aligning closely with their subsequent season trajectories.

Overall, the aggregate mid-season simulation results demonstrate that expected goals provide a meaningful early-warning signal for both positive and negative season outcomes.  These findings underscore the role of randomness in football and highlight the value of expected goals as a probabilistic baseline for inference and uncertainty quantification, rather than as a deterministic predictor of rare season outcomes.

% --- Figure 9 inline (no floating) ---
\renewcommand{\figurename}{Figure}
\begin{center}
\includegraphics[width=\linewidth]{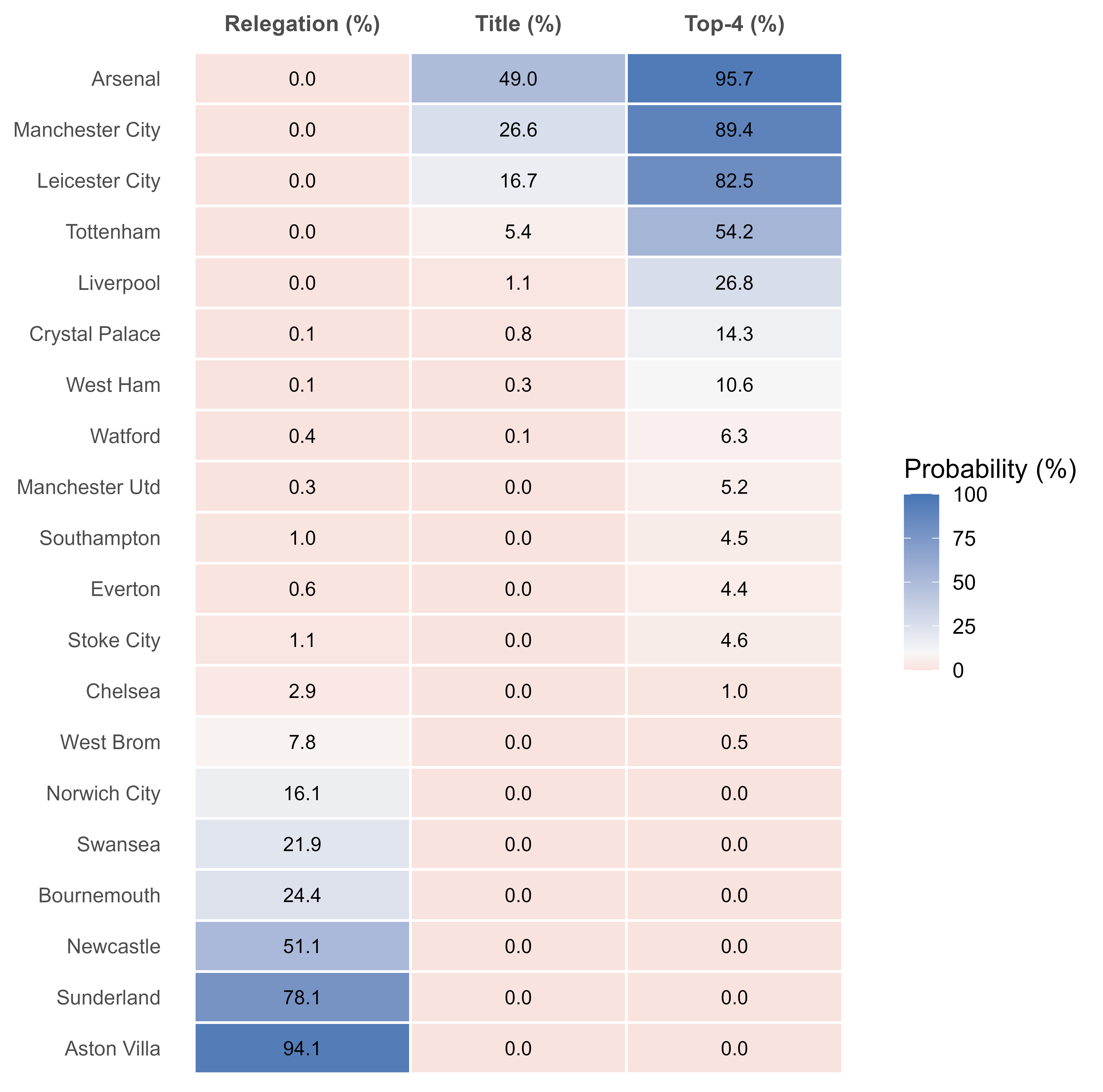} 
\captionof{figure}{Mid-season outcome probabilities based on xG simulations: title hopes, top-4 chances, and relegation risks}
\label{fig:pred_vs_real}
\end{center}

\subsubsection{Leicester City: A Tail Outcome Under Mid-Season Inference}

Leicester City’s 2015/16 championship represents the most prominent extreme realisation in our mid-season simulation framework. Conditioning exclusively on first-half expected goals, Leicester achieved an average simulated finishing position within the top four and was assigned a non-negligible probability of winning the title. While this probability remained substantially lower than that of traditional contenders such as Arsenal and Manchester City, it was strictly positive, reflecting Leicester’s strong underlying performance by mid-season.

Their eventual title win therefore constitutes a low-probability but feasible outcome under the ex ante predictive distribution implied by expected goals. 

\subsubsection{Relative Over- and Under-Performance at Mid-Season}

Beyond Leicester City, the mid-season simulations highlight several teams whose realised outcomes diverged meaningfully from their xG-based expectations. Manchester United and Chelsea, in particular, underperformed relative to their predicted second-half points and ranks, despite generating competitive first-half xG profiles. These deviations suggest inefficiencies in chance conversion, defensive execution, or contextual factors such as injuries and tactical instability that are not fully captured by xG alone.

By contrast, teams such as Arsenal and Manchester City exhibited realised outcomes broadly consistent with their mid-season expectations, reinforcing the model’s ability to distinguish genuine contenders from mid-table uncertainty. Overall, these patterns underscore the probabilistic nature of the framework expected goals constrain the range of likely outcomes and identify structural strengths, but do not preclude meaningful deviations driven by stochastic match events and executional variance.

\section{Discussion and Further Research}
\subsection {Interpretation of Simulation Results}

The mid-season xG-based inference framework demonstrates a strong ability to recover the broad structure of the league under partial information. Conditioning exclusively on first-half expected goals, the simulations correctly identify the principal contenders for top-four positions, as well as teams facing elevated relegation risk. This indicates that expected goals provide a meaningful early diagnostic signal of underlying team quality before final outcomes are realised.

At the same time, the results highlight the inherent limitations of xG when applied to rare or extreme outcomes. Leicester City, despite exhibiting a strong first-half xG profile and ranking among the leading teams at mid-season, was assigned a relatively modest title probability. Their eventual championship therefore represents a low-probability realisation from the upper tail of the simulated outcome distribution rather than a contradiction of the model’s expectations. This distinction is critical under mid-season information, Leicester’s success was unlikely, but not implausible.

More generally, deviations between simulated expectations and realised outcomes underscore the role of factors not explicitly captured by shot-based models. Persistent finishing efficiency, defensive organisation, tactical adaptation, squad rotation, and injury dynamics may all amplify or suppress realised performance relative to xG-based baselines. As such, expected goals should be interpreted as constraining the range of plausible outcomes rather than deterministically predicting final league positions.

Conversely, several teams traditionally viewed as title contenders most notably Arsenal and Manchester City retained the highest mid-season title probabilities, consistent with their strong underlying xG profiles. While their realised outcomes fell short of simulated expectations, these discrepancies further illustrate the influence of stochastic variation and contextual factors over the second half of the season.

Taken together, the mid-season results reinforce the central contribution of this study expected goals are most informative as an ex ante inference tool and early-warning signal, capable of identifying structural strength and quantifying uncertainty, while explicitly allowing for rare but consequential deviations that define football seasons.

\subsection{Model Limitations}
Despite the utility of xG modeling, several limitations must be acknowledged:
\begin{itemize}[label=\textbullet]

 \item Logistic regression treats all shots as independent events, ignoring temporal or tactical sequences (e.g., sustained pressure, momentum). While this assumption is standard in shot-based xG modeling, it may underrepresent clustering effects such as sustained pressure or momentum within matches.
 \item Due to data availability constraints, the model conditions on shot characteristics alone and does not explicitly incorporate defensive pressure, goalkeeper positioning, or team shape, all of which may influence conversion probability.
  \item The same xG value is assigned regardless of individual player finishing ability, potentially obscuring systematic differences. For example, between elite attackers and more defensively oriented players.
  \item Team-specific styles (e.g., compact defense, high pressing) are not incorporated, yet they significantly influence match outcomes.
 \item Poisson-based simulation captures the inherent randomness of goal scoring, but in small samples, it can exaggerate extreme outcomes. For example, a team with xG $\approx$ 1 may occasionally be simulated to score 4-5 goals, inflating variance and assigning non-negligible probabilities to improbable events. While these effects average out over thousands of runs, they partly explain why the model may understate favorites and overstate long shots. This behavior reinforces the interpretation of the simulations as probabilistic baselines rather than precise generators of individual season outcomes.

 \end{itemize}

\subsection {Further Work}

To enhance predictive power and model practicality, several improvements are needed, such as:
\begin{itemize}[label=\textbullet]
\item Incorporating player-specific scoring ability and finishing efficiency to personalize xG assignments.
\item Adding contextual features such as defensive pressure, number of defenders, and goalkeeper positioning to better reflect shot difficulty.
\item Employing sequence-aware models (e.g., RNNs or attention mechanisms) to capture momentum and build-up play.
\item Assigning weights for team style, pace, and pressing intensity to more accurately simulate league dynamics.
\item Leveraging advanced approaches such as Bayesian methods or ensemble learners to improve performance.
\item Extending the framework to other leagues (e.g., Serie A, La Liga) to assess model transferability.
\end{itemize}

\begin{appendices}

\section*{Appendix A: Ex Post Benchmark Using Full-Season xG}

For completeness, we report a retrospective benchmark in which the simulation framework is conditioned on full-season expected goals. Because this analysis uses information unavailable at mid-season, it does not constitute a feasible ex ante forecast. Instead, it provides an upper bound on achievable model fit, against which the mid-season inference results in Section~4 can be contrasted.

Table~6 reports rank- and point-level agreement between simulated and realized outcomes under this full-information setting.
\setcounter{table}{5}
\begin{table}[ht]
\centering
\caption{Ex post benchmark: full-season xG fit (upper bound)}
\label{tab:appendix_full_season_benchmark}
\begin{tabular}{llcc}
\toprule
\textbf{Metric} & \textbf{Definition} & \textbf{Points} & \textbf{Ranks} \\
\midrule
Spearman's $\rho$ & Rank correlation (monotonic association) & 0.837 & 0.824 \\
Pearson's $r$     & Linear correlation coefficient           & 0.871 & 0.862 \\ 
$R^2$             & Proportion of variance explained         & 0.758 & 0.744 \\[0.5em]
RMSE              & Root Mean Squared Error                  & 7.624 & 2.922 \\
MAE               & Mean Absolute Error                      & 6.102 & 2.220 \\
\bottomrule
\end{tabular}
\end{table}

\section*{Appendix B: Convergence of average team ranks with increasing simulations}
Figure 10 shows that the anticipated average ranks quickly come together as the number of simulations goes up. This shows that the expected outcomes are overall stable across teams.

\renewcommand{\thefigure}{10}
\setcounter{figure}{0} % reset figure counter

\begin{figure}[H]
    \centering
    \includegraphics[width=1.17\textwidth]{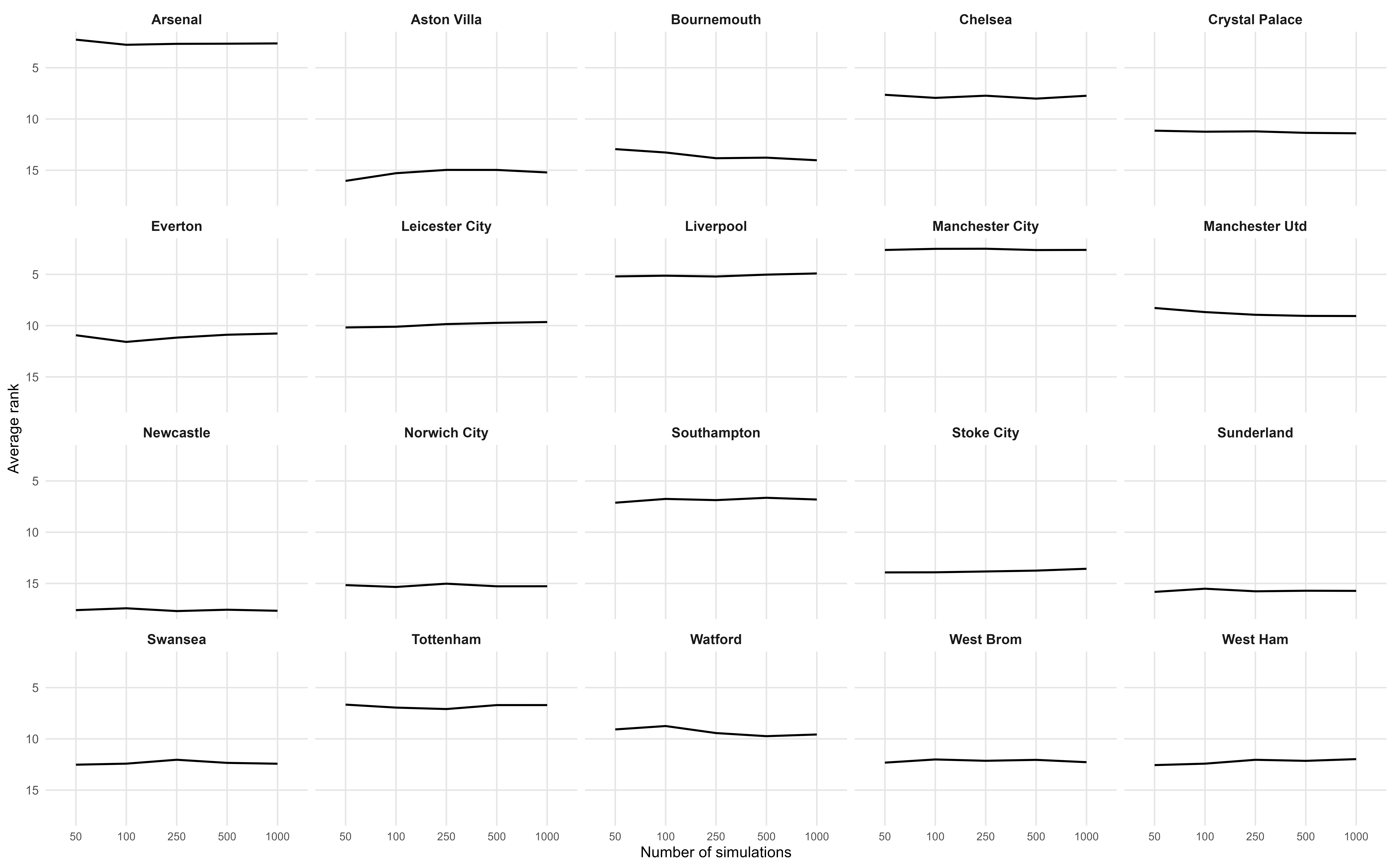}
    \caption{Convergence of average predicted ranks with increasing simulations}
    \label{fig:appendix1}
\end{figure}

\renewcommand{\thefigure}{11}
\section*{Appendix C: Leicester City: A Descriptive Illustration}
The figure 11 shows that Leicester accumulated points at a faster rate than underlying xG, indicating early overperformance relative to chance creation. This descriptive trajectory complements the league-level analysis by providing an intuitive illustration of Leicester’s divergence between realised outcomes and underlying performance.
\begin{figure}[H]
    \centering
    \includegraphics[width=\textwidth]{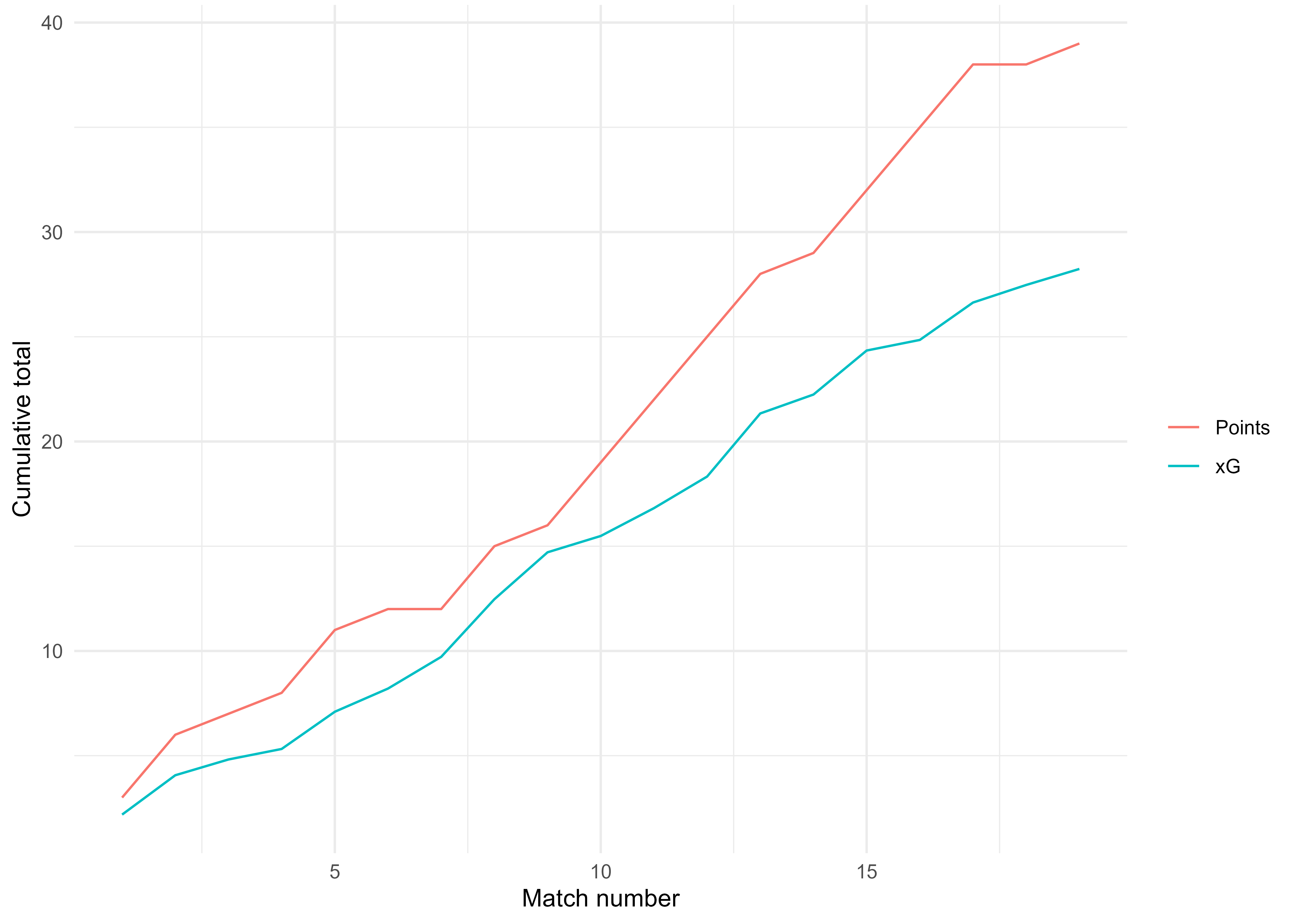}
    \caption{Commulative xG vs points (first half)}
    \label{fig:fig9}
\end{figure}
\end{appendices}

%\begin{acknowledgement}
  % This work has been supported by the project "SMARTsports": “Statistical Models and AlgoRiThms in sports. Applications in professional and amateur contexts, with able-bodied and disabled athletes”, funded by the MIUR Progetti di Ricerca di Rilevante Interaesse Nazionale (PRIN) Bando 2022 - grant n. 2022R74PLE (CUP J53D23003860006).
%\end{acknowledgement}

%%=============================================%%
%% For presentation purpose, we have included  %%
%% \bigskip command. Please ignore this.       %%
%%=============================================%%

\backmatter

%%\bmhead{Supplementary information}

%%If your article has accompanying supplementary file/s please state so here. 

%%Authors reporting data from electrophoretic gels and blots should supply the full unprocessed scans for key as part of their Supplementary information. This may be requested by the editorial team/s if it is missing.

%%Please refer to Journal-level guidance for any specific requirements.

\bmhead{Acknowledgements}

Language refinement for this manuscript was supported by Overleaf’s editing tools and Grammarly.

\section*{Data Availability}
The datasets supporting the conclusions of this article are available in the following public repositories: 
\begin{itemize}
    \item Kaggle: \url{https://www.kaggle.com/datasets/secareanualin/football-events}
    \item GitHub (StatsBomb Open Data): \url{https://github.com/statsbomb/open-data}
    \item StatsBomb Resource Centre: \url{https://statsbomb.com/resource-centre/}
    \item Hudl StatsBomb Platform: \url{https://www.hudl.com/en_gb/products/statsbomb}
\end{itemize}

\section*{Funding}
This work has been supported by the project "SMARTsports": “Statistical Models and AlgoRiThms in sports. Applications in professional and amateur contexts, with able-bodied and disabled athletes”, funded by the MIUR Progetti di Ricerca di Rilevante Interesse Nazionale (PRIN) Bando 2022 - grant n. 2022R74PLE (CUP J53D23003860006).

\section*{Competing Interests}
The authors declare no competing interests.






\section*{References}

\begin{list}{}{\setlength{\leftmargin}{0em}\setlength{\itemindent}{-1em}\setlength{\itemsep}{1ex}}
\item Anzer, G., and P. Bauer. 2021. 
“A Goal Scoring Probability Model for Shots Based on Synchronized Positional and Event Data in Football (Soccer).” 
\emph{Frontiers in Sports and Active Living} 3: 624475.

\item Bandara, I., S. Shelyag, S. Rajasegarar, D. Dwyer, E. J. Kim, and M. Angelova. 2024. 
“Predicting Goal Probabilities with Improved xG Models Using Event Sequences in Association Football.” 
\emph{PLOS ONE} 19 (10): e0312278.

\item Brechot, M., and R. Flepp. 2020. 
“Dealing with Randomness in Match Outcomes: How to Rethink Performance Evaluation in European Club Football Using Expected Goals.” 
\emph{Journal of Sports Economics} 21 (4): 335–62.

\item Cox, A., and C. Philippou. 2022. 
“Measuring the Resilience of English Premier League Clubs to Economic Recessions.” 
\emph{Soccer \& Society} 23 (4–5): 482–99.

\item Eggels, H., R. Van Elk, and M. Pechenizkiy. 2016. 
“Explaining Soccer Match Outcomes with Goal Scoring Opportunities Predictive Analytics.” 
In \emph{Proceedings of the 3rd Workshop on Machine Learning and Data Mining for Sports Analytics (MLSA 2016)}. Aachen: CEUR-WS.org.

\item Egidi, L., and N. Torelli. 2021. 
“Comparing Goal-Based and Result-Based Approaches in Modelling Football Outcomes.” 
\emph{Social Indicators Research} 156 (2): 801–13.

\item Elsharkawi, M., R. H. Ali, and T. A. Khan. 2025. 
“Crafting a Player Impact Metric through Analysis of Football Match Event Data.” 
\emph{Journal of Computational Mathematics and Data Science} 15: 100115.

\item Herbinet, C. 2018. 
“Predicting Football Results Using Machine Learning Techniques.” 
MEng thesis, Imperial College London.

\item Kharrat, T., I. G. McHale, and J. L. Peña. 2020. 
“Plus–Minus Player Ratings for Soccer.” 
\emph{European Journal of Operational Research} 283 (2): 726–36.

\item Kubayi, A. 2020. 
“Analysis of Goal Scoring Patterns in the 2018 FIFA World Cup.” 
\emph{Journal of Sports Sciences}.

\item Lago, C. 2009. 
“The Influence of Match Location, Quality of Opposition, and Match Status on Possession Strategies in Professional Association Football.” 
\emph{Journal of Sports Sciences} 27 (13): 1463–69.

\item Liu, H., M. Á. Gomez, C. Lago-Peñas, and J. Sampaio. 2015. 
“Match Statistics Related to Winning in the Group Stage of 2014 Brazil FIFA World Cup.” 
\emph{Journal of Sports Sciences} 33 (12): 1205–13.

\item Macdonald, B. 2012. 
“An Expected Goals Model for Evaluating NHL Teams and Players.” 
In \emph{Proceedings of the 2012 MIT Sloan Sports Analytics Conference}, March.

\item Macrì Demartino, R., L. Egidi, and N. Torelli. 2024. 
“Alternative Ranking Measures to Predict International Football Results.” 
\emph{Computational Statistics}, 1–19. https://doi.org/10.1007/s00180-024-XXXX.

\item Mead, J., A. O’Hare, and P. McMenemy. 2023. 
“Expected Goals in Football: Improving Model Performance and Demonstrating Value.” 
\emph{PLOS ONE} 18 (4): e0282295.

\item Nguyen, Q. 2021. 
“Poisson Modeling and Predicting English Premier League Goal Scoring.” 
\emph{arXiv Preprint} arXiv:2105.09881.

\item Nipoti, B., and L. Schiavon. 2025. 
“Expected Goals under a Bayesian Viewpoint: Uncertainty Quantification and Online Learning.” 
\emph{Journal of Quantitative Analysis in Sports} 21 (1): 37–50.

\item Rathke, A. 2017. 
“An Examination of Expected Goals and Shot Efficiency in Soccer.” 
\emph{Journal of Human Sport and Exercise} 12 (2): 514–29.

\item Secărean, A. 2025. 
\emph{Football Events} [Dataset]. Kaggle. Available at: https://www.kaggle.com/datasets/secareanualin/football-events (accessed February 9, 2025).

\item title, A. P., and M. Suguna. 2023. 
“Data-Driven Player Recruitment in Football.” 
In \emph{2023 2nd International Conference on Automation, Computing and Renewable Systems (ICACRS)}, 844–50. IEEE.

\item Souza, N. J. D., H. N. Samrudh, S. Gautham, B. U. Shaman Bhat, and N. Nagarathna. 2021. 
“Football Game Analysis and Prediction.” 
In \emph{Soft Computing and Signal Processing: Proceedings of 3rd ICSCSP 2020, Volume 1}, 167–80. Singapore: Springer Singapore.

\item Skripnikov, A. V., A. Cemek, and D. Gillman. 2025.  
“Leveraging Minute-by-Minute Soccer Match Event Data to Adjust Team’s Offensive Production for Game Context.”  
\emph{Journal of Quantitative Analysis in Sports}. https://doi.org/10.1515/jqas-2024-0162.

\item Spearman, W. 2018. 
“Beyond Expected Goals.” 
In \emph{Proceedings of the 12th MIT Sloan Sports Analytics Conference}, February, 1–17.

\item Umami, I., D. H. Gautama, and H. R. Hatta. 2021. 
“Implementing the Expected Goal (xG) Model to Predict Scores in Soccer Matches.” 
\emph{International Journal of Informatics and Information Systems} 4 (1): 38–54.

\item Vilela, J. A. T. F. 2024. 
“Exploring Stochastic Efficiency Analysis for Expected Goals in Football: Assessing Offensive Efficiency Across Europe’s Major Football Leagues.” 
Master’s thesis, Universidade NOVA de Lisboa (Portugal).

\item Wang, S. H., Y. Qin, Y. Jia, and I. K. E. Igor. 2022. 
“A Systematic Review about the Performance Indicators Related to Ball Possession.” 
\emph{PLOS ONE} 17 (3): e0265540.

\item Wheatcroft, E. 2021. 
“Forecasting Football Matches by Predicting Match Statistics.” 
\emph{Journal of Sports Analytics} 7 (2): 77–97.

\end{list}

%\bibliographystyle{...}
%\bibliography{...}
\end{document}